\documentstyle [aps,twocolumn,psfig]{revtex}
\catcode`\@=11
\newcommand{\be}{\begin{equation}}
\newcommand{\ee}{\end{equation}}
\def\pmb#1{\setbox0=\hbox{#1}
\kern-.025em\copy0\kern-\wd0
\kern.05em\copy0\kern-\wd0
\kern-.025em\raise.0433em\box0}

\begin{document}
\title{Heisenberg frustrated magnets: a nonperturbative approach}
\author{M. Tissier, B. Delamotte, D. Mouhanna}
 
\vspace{0.5cm}
 
\address{Laboratoire de Physique Th\'eorique et Hautes Energies. Universit\'es Paris
 VI-Pierre et Marie Curie - Paris VII-Denis Diderot, 2 Place Jussieu, 75252 Paris Cedex
 05, France.}

\vspace{3cm}

\address{\mbox{ }}
\address{\parbox{14cm}{{\rm Frustrated magnets are a notorious example where the usual
 perturbative methods are in conflict. Using a nonperturbative Wilson-like approach, we
 get a coherent picture of the physics of Heisenberg frustrated magnets everywhere between
 $d=2$ and $d=4$. We recover all known perturbative results in a single framework and find
 the transition to be weakly first order in $d=3$. We compute effective exponents in good
 agreement with numerical and experimental data. 
}}}
\address{\mbox{ }}
\address{\parbox{14cm}{\rm PACS No: 75.10.Hk, 64.60.-i,  11.10.Hi, 11.15.Tk}}

\maketitle
\makeatletter
\global\@specialpagefalse
\makeatother
\vspace{0.5cm}

Understanding the effect of competing interactions in three dimensional classical spin
 systems is one of the great challenges of condensed matter physics. However, after twenty
 five years of investigations, the nature of the universality class for the phase
 transition of the simplest frustrated model, the antiferromagnetic Heisenberg model on a
 triangular lattice (AFHT model), is still a strongly debated
 question$^{\cite{kawamura10}}$. Due to frustration, the ground state of the AFHT model is
 given by a canted  configuration -- the famous 120$^{\circ}$ structure -- that implies a
 matrix-like order parameter$^{\cite{kawamura2}}$ and thus, the possibility of a new
 universality class.  Experiments performed on materials supposed to belong to the AFHT
 universality class display  indeed  exponents different from those of the standard $O(N)$
 universality class: for VCl$_2$$^{\cite{kadowaki}}$: $\beta=0.20(2),\gamma=1.05(3),
 \nu=0.62(5)$, for  VBr$_2$$^{\cite{wosnitza}}$: $\alpha=0.30(5)$, for 
 CuFUD$^{\cite{koyama}}$: $\beta=0.22(2)$ and for 
 Fe[S$_2$CN(C$_2$H$_5$)$_2$]$_2$Cl$^{\cite{defotis1,defotis2,defotis3}}$: $\beta=0.24(1),
 \gamma=1.16(3)$. These results however call for several comments.  First, the exponents
 violate  the scaling relations, at least by two standard deviations. Second, they differ
 significantly from those obtained by Monte Carlo (MC) simulations performed either
 directly on the AFHT model ($\nu\simeq 0.59(1),\gamma\simeq 1.17(2),\beta\simeq
 0.29(1),\alpha \simeq 0.24(2)$), and on models supposed to belong to the same
 universality class: AFHT with rigid constraints ($\nu= 0.504(10),\gamma= 1.074(29),\beta=
 0.221(9),\alpha = 0.488(30) $), dihedral (i.e. $V_{3,2}$ Stiefel) models ($\nu\simeq
 0.51(1),\gamma\simeq 1.13(2),\beta\simeq 0.193(4),\alpha \simeq 0.47(3) $). See
 Ref.{\cite{loison2}} for a review, and references therein. Finally, the anomalous
 dimensions $\eta$ obtained by means of scaling relations is found to be negative in
 experiments as well as in MC simulations, a result forbidden by first principles for
 second order phase transitions$^{\cite{zinn}}$. All these results are hardly compatible
 with the assumption of universality. It has been proposed that  the exponents  are, in
 fact, effective exponents characterizing a very weakly first order transition, the
 so-called ``almost second order phase transition$^{\cite{zumbach7,zumbach,zumbach4}}$''.

From the theoretical point of view the situation is also very unsatisfactory since one
 does not have a coherent picture of the expected critical behaviour of the AFHT model
 between two and four dimensions. On the one hand, the weak coupling expansion performed
 on the suitable Landau-Ginzburg-Wilson (LGW) model in the vicinity of $d=4$ leads  to a
 first order phase transition due to the lack of a stable fixed
 point$^{\cite{bailin,garel,yosefin}}$. On the other hand, the low temperature expansion
 performed around two dimensions on the Non-Linear Sigma (NL$\sigma$) model predicts a
 second order phase transition of the standard $O(4)/O(3)$ universality
 class$^{\cite{aza4}}$. Since there is no indication that these perturbative results
 should fail in their respective domain of validity -- i.e. for small $\epsilon=4-d$ and
 small $\epsilon=d-2$ -- this situation raises two problems. First, and contrary to what
 happens in the non-frustrated case, one cannot safely predict the three dimensional
 behaviour from na\"{\i}ve extrapolations of the perturbative results. Although a direct
 computation in three dimensions, possible on the LGW model$^{\cite{antonenko3,loison1}}$,
 can circumvent this difficulty, such an approach  misses a second fondamental problem:
 the incompatibility between the symmetries of the NL$\sigma$ and LGW models. Indeed, the
 renormalization group flow drives the NL$\sigma$ model action towards an $O(4)$
 symmetric regime, more symmetric than the microscopical system, a phenomenon that {\it
 cannot} occur within all previous treatments of the LGW model (see ref. {\cite{aza4}} and
 below). The LGW model is therefore unable to find the $O(4)$ behaviour which has been
 nevertheless observed numerically in $d=2$$^{\cite{southern}}$. This raises serious
 doubts on the perturbative analysis of the LGW model away from $d=4$. Reciprocally, the
 {\it perturbative} analysis of the NL$\sigma$ model, based on a Goldstone mode expansion,
  predicts an $O(4)/O(3)$ fixed point everywhere between $d=2$ and $d=4$, as for the $N=4$
 ferromagnetic model, in contradiction with the perturbative LGW results and the
 experimental and numerical situation in $d=3$. All this suggests that non perturbative
 features could play a major role and thus imposes to go beyond the standard perturbative
 approaches.

In this letter we realize this program by using the Wilson renormalization group
 framework$^{\cite{wilson}}$. We obtain a coherent picture of the physics of the AFHT
 model everywhere between $d=2$ and $d=4$. We find that the fixed point expected from the
 NL$\sigma$ model approach exists indeed in the vicinity of $d=2$ but disappears below --
 and close to -- three dimensions. The transition for AFHT in $d=3$ is thus {\it weakly}
 first order contrary to the different predictions of both a new universality
 class$^{\cite{kawamura2}}$ and an $O(4)/O(3)$ second order behaviour$^{\cite{aza4}}$. We
 get effective exponents compatible with the numerical and experimental data quoted above.
 For generalization to $N>4$-component spins, we find the transition in $d=3$ to be second
 order with exponents in good agreement with recent extensive MC simulations -- contrary to those found from three loop Pad\'e-Borel resummed series$^{\cite{loison1}}$.

Our approach relies on the concept of effective average
 action$^{\cite{wetterich2,wetterich3}}$, $\Gamma_k[\phi]$, which is a coarse grained free
 energy where only fluctuations with momenta $q\ge k$ have been integrated out. The field
 $\phi$ corresponds to an average order parameter at scale $k$, the analog of a
 magnetization at this scale. At the scale of the inverse lattice spacing $\Lambda$,
 $\Gamma_{k=\Lambda}$ is the continuum limit of the lattice hamiltonian obtained, for
 example, by means of an Hubbard-Stratonovich transformation. On the other hand, the usual
 free energy $\Gamma$, generating one particle-irreducible correlation functions,  is 
 recovered in the limit $k\to 0$. 
The $k$-dependence of $\Gamma_k$ is controlled by an exact evolution
 equation$^{\cite{wetterich1,morris1}}$:
\begin{equation}
{\partial \Gamma_k\over \partial t}={1\over 2} \hbox{Tr} \left\{(\Gamma_k^{(2)}+R_k)^{-1}
 {\partial R_k\over \partial t}\right\}
\label{renorm}
\end{equation}
where $t=\ln \displaystyle {k / \Lambda}$. The trace has to be understood as a momenta
 integral as well as a summation over internal indices. In Eq.(\ref{renorm}), $R_k$ is the
 effective infrared cut-off which suppresses the propagation of modes with momenta $q<k$.
 A convenient cut-off is provided by$^{\cite{wetterich1,morris4}}$: $R_k(q)=Z
 q^2/(\exp(q^2/k^2)-1)$, where $Z$ is the field renormalization. In Eq.(\ref{renorm}),
 $\Gamma_k^{(2)}$ is  the {\it exact field-dependent} inverse propagator -- i.e. the
 second derivative of $\Gamma_k$.

The effective average action $\Gamma_k$ is a  functional  invariant under the symmetry
 group of the system and thus depends on all the invariants  built from the average order
 parameter. In our case, it is well known that the order parameter is a set of
two vectors $\vec{\phi_1}$ and $\vec{\phi_2}$ that can be gathered in a real $N\times 2$
 matrix  $\phi_{ab}$ for $N$-component spins$^{\cite{kawamura2}}$. The symmetry of the
 system is the usual spatial rotation group $O(N)$ times a $O(2)$ corresponding to the
 symmetry of the underlying triangular lattice$^{\cite{aza4}}$. This $O(2)$  is realized
 on $\phi_{ab}$ as a right $O(2)$ ``rotation" that  turns the $\vec{\phi_i}$ into each
 other. There are two independent $O(N)\otimes O(2)$ invariants built out of $\phi_{ab}$:
 $\rho={\hbox{Tr}}\ ^{t}\phi\phi$ and $\tau={1\over 2}{\hbox{Tr}} (^{t}\phi\phi)^2-{1\over
 4}({\hbox{Tr}}\ ^{t}\phi\phi)^2$. 

The exact effective average action involves all the powers of $\rho,\tau$ and of
 derivative terms, and so  Eq.(\ref{renorm}) is a nonlinear functional  equation,  too
 difficult to be
 solved exactly in general. We therefore need  to truncate it. One possibility is to keep
 in $\Gamma_k$ only the momentum (i.e. derivative)-independent part, an approximation
 called the Local Potential Approximation (LPA). In the case of frustrated magnets, this
 has been considered by Zumbach$^{\cite{zumbach7,zumbach,zumbach4}}$. This approximation
 however misses the field-renormalization and worse, as described below, the phenomenon of
 enlarged symmetry around $d=2$ found perturbatively in the NL$\sigma$
 model$^{\cite{aza4}}$. This does not mean that this approximation is not useful: it is
 simply, in essence, unable to answer the question of the matching of the different
 perturbative approaches. Another truncation is however possible which preserves this
 possibility:  it consists in an expansion of $\Gamma_k$ around its minimum in order to
 keep a finite number of monomials in the invariants $\rho$ and $\tau$ while including the
 derivative terms allowing to recover the different perturbative results. We choose the
 simplest such truncation:
\begin{equation}
\begin{array}{l}
\Gamma_k= \displaystyle \int d^dx \left\{{Z\over 2}  \nabla\phi_{ab}\nabla \phi_{ab}+ {
 \omega\over 4}\ (\epsilon_{ab}\phi_{ca} \nabla \phi_{cb})^2 \right. \nonumber
\\
\\
\left. \displaystyle\hskip1.7cm+{\lambda\over 4}\left({\rho\over
 2}-\kappa\right)^2+{\mu\over 4}\tau\right\}
\label{action}
\end{array}
\end{equation}
where $\left\{\omega, \lambda, \kappa, \mu,Z\right\}$ are the coupling constants which
 parametrize the model. All terms but one - the ``current term'' $(\epsilon_{ab}\phi_{ca}
 \nabla \phi_{cb})^2$ -  are very natural and correspond to those appearing in the usual
 LGW action that realizes the symmetry breaking scheme of frustrated magnets. Indeed for
 $\lambda$ and $\mu \ge 0$, the minimum of the action is realized by a configuration of
 the form $\phi_{ab}^{min}=\sqrt{\kappa} R_{ab}$, where $R_{ab}$ is a matrix built with
 two orthonormal $N$-component vectors. The symmetry of this minimum is a product of a
 diagonal $O(2)$ group and a residual $O(N-2)$ group. The symmetry breaking scheme is thus
 $O(N)\otimes O(2)\to O(N-2)\otimes O(2)_{diag}$$^{\cite{aza4}}$. Note that for
 $\phi_{ab}=\phi_{ab}^{min}$ one has: $\rho=2\kappa$ and $\tau=0$ so that Eq.(\ref{action}) corresponds indeed to a quartic expansion around the minimum. The
 spectrum in the low temperature phase consists in $2N-3$ Goldstone modes and three
 massive modes: one singlet of mass $m_1=\kappa \lambda$ and  one doublet of mass
 $m_2=\kappa\mu$ which correspond to fluctuations of the relative angle and  of the norms
 of the two vectors $\vec{\phi_1}$ and $\vec{\phi_2}$.

Without the current term, the truncation Eq.(\ref{action}) is however not sufficient in
 our case. This term plays a crucial role since, for $N=3$, it allows the model to enlarge
 its symmetry from $O(3)\otimes O(2)$ to $O(3)\otimes O(3)\sim O(4)$ at the fixed point
 around $d=2$, leading to the well known $O(4)/O(3)$ behaviour$^{\cite{aza4}}$. The
 current term is systematically discarded in the perturbative treatment of the LGW model
 around four dimensions, for the - correct - reason that it is power-counting irrelevant.
 Here we can include it in our {\it ansatz} since it is anyway present in the full
 effective action $\Gamma_k$ and, in fact, {\it must} include it since it becomes relevant
 somewhere between two and four dimensions. The formalism we use is in charge to decide
 where it is important.

Let us emphasize that the effective average action method leads to non trivial and/or new
 results even within a quartic truncation of $\Gamma_k$. One can mention the
 Kosterlitz-Thouless phase transition$^{\cite{grater}}$, low energy
 QCD$^{\cite{jungnickel1}}$, the abelian Higgs model and
 superconductivity$^{\cite{bergerhoff1,bergerhoff2}}$, etc. The accuracy of the results
 thus obtained depends on two main features: i) the smallness of the anomalous dimension
 $\eta$ and ii) the fact that the thermodynamics of the system is controlled by a unique
 minimum of $\Gamma_k$. Note finally that this technique has been successfully employed in
 the case of the principal chiral model to solve a conflict between perturbative
 approaches$^{\cite{tissier1}}$, similar to what is studied here. However we stress that
 in the principal chiral case, there was no conflict between the symmetries of the LGW and
 NL$\sigma$ models. 

The flow equations for the different coupling constants $\kappa$, $\lambda$, $\mu$,
 $\omega$ and  $Z$ are derived by using Eq.(\ref{renorm}) and Eq.(\ref{action}) along
 the same lines as in \cite{jungnickel1}. The explicit recursion equations are too long to
 display and not particularly illuminating (see {\cite{site}}). Moreover, they require a
 numerical analysis, apart in $d=2+\epsilon$ and in $d=4-\epsilon$ where, as we now see,
 they get analytically tractable.

{\sl The physics around two dimensions}. Around two dimensions, one expects that the
 perturbative  ``Goldstone mode" expansion of the NL$\sigma$ model works well.  In  the
 Goldstone regime, the fluctuations of the modulus of ${\vec\phi}_1$ and ${\vec\phi}_2$
 and of their relative angle are frozen. This corresponds to the large mass limit $m_{1r}$,
 $m_{2r}\to \infty$. In this limit, our equations greatly simplify since the coupling
 constants divide in two sets $\lbrace \kappa, \omega, Z\rbrace$ and $\lbrace \lambda, \mu
 \rbrace$ that do not mix. We only quote here the flow equations for the renormalized
 coupling constants of the first set:
\begin{equation}
\left\{
\begin{array}{l}
\displaystyle{d\kappa_r\over dt}=-(d-2+\eta)\kappa_r +{N-2\over 2\pi} +{1\over 4\pi (1+
 \kappa_r \omega_r)}
\\
\displaystyle{{d\omega_r\over dt}=(-2 + d + 2 \eta) \omega_r +}
\\
\ \ \ \ \displaystyle{1+\kappa_r \omega_r +
(N-1)\kappa_r^2 \omega_r^2 +(N-2)\kappa_r^3 \omega_r^3\over 2 \kappa_r^2 \pi(1+ \kappa_r
 \omega_r)}\ \\
\\
\displaystyle\eta=-{d \ln Z \over dt}={3 + 4\kappa_r \omega_r + 2\kappa_r^2 \omega_r^2
 \over 4\kappa_r\pi(1 + \kappa_r \omega_r)}
\end{array}
\right.
\label{perturb2d}
\end{equation}
These equations admit a fixed point for any $N>2$ of coordinates $\kappa_r \simeq
 1/\epsilon$, $\omega_r \simeq \epsilon$, while $\lambda_r,\mu_r \simeq$ cst. The masses
 $m_{1r}^*$, $m_{2r}^*$ are thus very large, proving the consistency of the limit.
In fact, modulo the change of variables: $\eta_1=\kappa_r$ and
$\eta_2= 2 \kappa_r (1+ \kappa_r \omega_r)$ the equations for $\kappa_r$ and $\omega_r$
 are exactly those obtained at one-loop in the perturbative analysis of the NL$\sigma$
 model$^{\cite{aza4}}$. For $N=3$, they admit a fixed point for which the model is
 $O(4)$-symmetric.  

Let us now recall how this phenomenon of enlarged symmetry for $N=3$ can be understood
 directly on the partition function. At the fixed point, the potential gets infinitely
 deep so that one recovers the hard constraints of the NL$\sigma$ model:
 ${\vec\phi_{1}}\perp{\vec\phi_{2}}$, and ${\vec\phi_{1}}^2 = {\vec\phi_{2}}^2=
 \kappa^*_r$. For $N=3$, this allows us to rewrite the current term as the kinetic term of
 a third vector, the cross product of the two others: $(\epsilon_{ab}\phi_{ca} \nabla
 \phi_{cb})^2\propto(\nabla {\vec\phi}_3)^2$ with ${\vec\phi}_3={\vec\phi}_1\wedge
 {\vec\phi}_2$. The order parameter  of the system  is then a trihedral of orthogonal
 vectors $({\vec\phi}_1,{\vec\phi}_2,{\vec\phi}_3)$. Thus  contrary to what could be
 expected from a na\"{\i}ve expansion in powers of the fields, the current term plays a
 role as important as the usual kinetic terms. At the fixed point, $\omega_r$ takes a
 value such that the three vectors play a symmetric role and the symmetry breaking scheme
 is  $O(3)\otimes O(3)/ O(3)\sim O(4)/O(3)$ instead of $O(3)\otimes O(2)/ O(2)$. Such a
 result is of course missed within the LPA$^{\cite{zumbach7,zumbach,zumbach4}}$.
 Therefore, the presence of the current term does not only improve the accuracy of the
 calculation, it is necessary for its consistency. 

{\sl The physics around four dimensions.} Around four dimensions, we have expanded our
 equations at leading order in the coupling constants $\lambda_r$ and $\mu_r$. At this
 order the current term decouples and we are left with the following equations for the
 quartic coupling constants:
\begin{equation}
\left\{
\begin{array}{l}
\displaystyle{d\lambda_r\over dt}= (-4 + d)\lambda_r + {1\over 16\pi^2}(4 \lambda_r \mu_r
 + 4\mu_r^2 + \lambda_r^2 (4 + N))\\
\\
\displaystyle{d\mu_r\over dt}=(-4 + d )\mu_r + {1\over 16 \pi^2}(6\lambda_r\mu_r + N
 \mu_r^2).
\end{array}
\right.
\end{equation}
They are those obtained at one loop from the LGW approach$^{\cite{bailin}}$. These flow
 equations admit a stable fixed point for $N>N_c\simeq21.8$, attesting that the phase
 transition is second order. For $N<N_c$ the transition is first order since no fixed
 point is found.

To higher orders, $N_c$ depends on the dimension. In $d=3$, three loop calculations
 resummed {\it \`a la} Pad\'e-Borel predict $N_c(d=3)=3.91$$^{\cite{antonenko3}}$. Note
 however that this calculation exhibits unusual behaviours compared to the $O(N)$ case:
 the coefficients of the series do not decrease monotonically and the series themselves
 are not alternate$^{\cite{loison1}}$. These features reveal the poor summability of the
 series. Finally, in the $N=6$ case, for which the transition is second order, the
 predictions based on a Pad\'e-Borel resummation, which provides $\nu=0.575$ and
 $\gamma=1.121$$^{\cite{loison1}}$, are in clear disagreement with recent numerical
 simulations, for which $\nu=0.700(11)$ and $\gamma=1.383(36)$$^{\cite{loison1}}$.

From this point of view our approach has several avantages: first, since it matches with
 the one loop perturbative results in $d=2$ and $d=4$ it is likely that the error does not
 vary much with the dimension -- a fact that has been confirmed in the $O(N)$ case for
 which the precision for a given truncation is almost uniform with $d$. Second, it does
 not rely on a Pad\'e-Borel resummation and therefore is free of the above mentionned
 problems of convergence.  Of course, our results will change while improving the {\it ansatz} Eq.({\ref{action}}) by incorporating terms of higher order in fields and
 derivatives. However, all cases already treated within the average action method suggest
 that the lowest order approximation gives fairly good results, even with this crude
 approximation. For example, in the ferromagnetic $O(3)$ model, one finds
 $\nu=0.703$$^{\cite{tetradis1}}$  which has to be compared to the six loop resummed
 perturbation series in three dimensions which provide  $\nu=0.705$$^{\cite{zinn}}$.

{\sl The physics between two and four dimensions.} Let us first study the fate of the
 fixed point found analytically in $d=2+\epsilon$ for $N=3$. By numerically integrating
 the flow equations, we find that this stable $O(4)/O(3)$ fixed point  describes a smooth
 trajectory in the coupling constant space while $d$ is increased. Our flow equations
 actually admit another -- but unstable -- fixed point, which moves toward the stable
 fixed point as the dimension is increased. At a critical dimension $d_c\simeq 2.87$, the
 two fixed points collapse and  disappear. Above $d_c$, no other stable fixed point is
 found and we conclude that the transition is first order in $d=3$. We thus show that the
 $O(4)/O(3)$ fixed point obtained from the NL$\sigma$ model plays no role in the three
 dimensional physics of frustrated magnets, as conjectured for example by Jolic\oe ur and
 David$^{\cite{jolicoeur4}}$ and Dobry and Diep$^{\cite{dobry}}$. We also discard the
 possibility of a new universality class conjectured on the basis of a na\"{\i}ve
 extrapolation of the $\epsilon=4-d$ calculation$^{\cite{kawamura10,kawamura2}}$. The
 proximity of $d_c$ with $d=3$ however let open the possibility of a very weakly first
 order phase transition with effective critical exponents. This behaviour manifests itself
 in our equations by the existence of a minimum  around which the RG flow slows down. This
 characterizes a very large, although finite correlation length $\xi$. A rough estimate of
 this correlation length -- a few hundred lattice spacings -- indicates that a
 pseudo-scaling behaviour can be observed although $\xi$ is not large enough to ensure a
 true universality. This could explain the broad spectrum of effective critical exponents
 found in experiments and numerical simulations. Although the flow equations do not have a
 fixed point, we are able to compute effective exponents by linearizing the flow equations
 around the minimum. We recover here the phenomenon of ``almost second order phase
 transition'' first introduced by Zumbach$^{\cite{zumbach7,zumbach,zumbach4}}$ within the
 LPA. To get accurate results we have to take into account the $\phi^6$-like terms in our
 {\it ansatz}. We find: $\nu=0.53$, $\gamma=1.03$ and $\beta=0.28$, which lie in between
 the various sets of exponents found experimentally and numerically (see above). For
 comparison Zumbach found $\nu\simeq0.63$ in the LPA$^{\cite{zumbach7,zumbach,zumbach4}}$,
 the difference being mainly due to
 the anomalous dimension.

Finally, we find a true fixed point in $d=3$ for $N$ larger than a critical value
 $N_c(d=3)\simeq 4$. For $N=6$, we get $\nu=0.74$ and $\gamma=1.45$ which compare well
 with the Monte Carlo data $\nu=0.700(11)$ and $\gamma=1.383(36)$$^{\cite{loison1}}$. They
 are close to the LPA results, where $\nu=0.76$$^{\cite{zumbach}}$, and much better than
 those obtained by a three-loop calculation in $d=3$$^{\cite{loison1}}$ (see
 above). We have checked that our exponents do not vary significantly when monomials of
 order six in the fields are included in the {\it ansatz} Eq.($\ref{action}$). 

To conclude, using a non perturbative method, we have reached a global understanding of
 frustrated Heisenberg magnets including a matching between previous perturbative
 predictions and a good agreement with experimental and numerical data. It remains to
 understand the very origin of the disappearance of the NL$\sigma$ model fixed point. 
 The role of non trivial topological configurations can be invoked. One can hope a
 complete understanding of this point through the average action method
 which successfully describes the Kosterlitz-Thouless phase transition$^{\cite{grater}}$. 

We thank J. Vidal for a careful reading of the manuscript

LPTHE is a laboratoire associ\'e au CNRS UMR 7589.
e-mail: tissier,delamotte,mouhanna@lpthe.jussieu.fr

\end{document}